\documentclass[fleqn,10pt]{wlscirep}
\usepackage[utf8]{inputenc}
\usepackage[T1]{fontenc}
\title{4.2\,K Sensitivity-Tunable Radio Frequency Reflectometry of a Physically Defined \textcolor{black}{P-}channel Silicon Quantum Dot}

\author[1,*]{Sinan Bugu}
\author[1]{Shimpei Nishiyama}
\author[2]{Kimihiko Kato}
\author[2]{Yongxun Liu}
\author[2]{Shigenori Murakami}
\author[2]{Takahiro Mori}
\author[3]{Thierry Ferrus}
\author[1,+]{Tetsuo Kodera}

\affil[1]{Tokyo Institute of Technology 2-12-1 Ookayama, Meguro-ku, Tokyo 152-8552, Japan}
\affil[2]{Device Technology Research Institute (D-Tech), National Institute of Advanced Industrial Science and Technology (AIST), Central 2, 1-1-1 Umezono, Tsukuba, Ibaraki, 305-8568, Japan}
\affil[3]{Hitachi Cambridge Laboratory, J. J. Thomson Avenue, CB3 0HE, Cambridge, United Kingdom}

\affil[*]{sinanbugu@gmail.com}

\affil[+]{kodera.t.ac@m.titech.ac.jp}

\begin{abstract}
	We demonstrate the measurement of p-channel silicon-on-insulator quantum dots at liquid helium temperatures by using a radio frequency (rf) reflectometry circuit comprising of two independently tunable GaAs varactors. This arrangement allows observing Coulomb diamonds at 4.2\,K under nearly best matching condition and optimal signal-to-noise ratio.  We also discuss the rf leakage induced by the presence of the large top gate in MOS nanostructures and its consequence on the efficiency of rf-reflectometry. These results open the way to fast and sensitive readout in multi-gate architectures, including multi qubit platforms.
\end{abstract}
\begin{document}
	
	\flushbottom
	\maketitle
	%
	%
	\thispagestyle{empty}

	\section*{Introduction}
	
	Hole spins confined in silicon structures have attracted considerable attention due to their ability to provide quantum bit architectures a longer coherence \cite{Veldhorst_2014} and faster gate operation times \cite{Hendrickx2020} compared to electrons in III-V materials \cite{Pribiag2013}. They also take advantage of an all electric field control due to the increase in the spin-orbit coupling strength \cite{voisin2016electrical,li2015pauli,maurand2016cmos, Szumniak2012}, a charge offset stability \cite{zimmerman2001excellent} necessary for scalability purposes as well as a necessary compatibility with current industry production lines in view of commercialisation. \cite{veldhorst2017silicon, vandersypen2017interfacing,gonzalez2020scaling}. However, to take full advantage of these properties for quantum computing applications, one has to circumvent the limitation in bandwidth resulting from the formation of the low-pass filter circuit made from the large input  impedance of the device or the measurement unit and the large cabling capacitance. In the first case, traditional transimpedance amplifiers are structurally limited to bandwidths of a few kilohertz due to a large input resistance and capacitance \cite{manoharan2008silicon}. Also one has to manage the intrinsic trade-off between output gain and bandwidth, with III-V amplifiers giving the highest speed while silicon based amplifiers generally providing a better gain. For the second case, the restriction lies in the length of cable that, all quality considered, has a capacitance in excess of 70\,pF/m \cite{gautschipiezoelectric}. Finally, device geometry and structure play a significant part in the reduction in bandwidth. For example, tunnel barrier-based sensors require resistances of a few h/e$^2$ for the observation of high quality Coulomb blockade oscillations that are used for monitoring of the charge state or for spin-to-charge conversion \cite{Elzerman_2003_PRB}.
	
	Reflectometry techniques in the radio frequency (rf) range have purposely been developed to address the former issue by incorporating the device into an LC resonant circuit and measuring the modulated reflected signal at a frequency close to the resonant condition \cite{Schoelkopf_1998}. In quantum dot architectures, such a circuit can be integrated either at the sensor, for example a single electron transistor or a quantum point contact close to the device to measure \cite{qin2006radio,cassidy2007single,reilly2007fast,noiri2020radio}, or directly at one of the control gate of the device itself \cite{colless2013dispersive,petersson2010charge,crippa2019gate,schroer2012radio,urdampilleta2015charge,gonzalez2015probing,Imtiaz_2018, schaal2020fast}. Both techniques have proven to be reliable with bandwidths in excess of 10\,MHz and are particularly suited to quantum dots where the low-stray parasitic capacitance, generally lower than 1\,pF, allows the use of moderate inductance sizes and values to obtain a high resonance frequency while retaining good impedance matching. Gated devices including double gate stacks, whose the top gate is generally large as it is used to create the inversion layer at the Si-SiO$_2$ interface and provide conduction to the device whereas the underlying set of gates only aim at creating tunnel barriers and adjusting the tunneling rates, \cite{nakagawa2003primary} provide a significant advantage in terms of charge noise screening, uniformity of the 2DEG compared with similar but doped structures. However, such advantages can be overcome by the formation of a low impedance leakage pathway to the ground for the rf signal that results from the large parasitic capacitance between the accumulation gate and the two-dimensional electron gas (2DEG) \cite{connors2020rapid,liu2020radio,noiri2020radio}. This is particularly severe for the area larger than a few tens of $\mbox{$\mu$m}^2 \hspace{0.1cm} $ \cite{bugu2021rf}.
	
	Aside bandwidth considerations, the reduction in the readout integration time has important implications. It allows improving readout fidelity once reaching timescales typically shorter than the relaxation time of the system $T_1$ and envisioning surface code-based applications when measurements become much faster than the coherence time $T_2$. However, shorter integration times unequivocally implies a reduction in the signal-to-noise ratio (SNR) \cite{Ibberson2020}. In the low-signal regime, when the changes in the ratio of resistance to capacitance are small, the SNR depends on the coupling coefficient \cite{ares2016sensitive, ibberson2019low} that becomes maximum at perfect matching condition. In order to achieve a high sensitivity while keeping perfect impedance matching, voltage-controlled capacitors, or varactors , are generally incorporated into the resonant circuit. Such an implementation has already been used to tune the resonant frequency of a quantum point contact in-situ at 1.8\,K \cite{muller2010situ}. Similarly, a couple of varactors have been used for impedance matching purposes to measure both the charge sensitivity and the impedance of a quantum dot at 1\,K \cite{ares2016sensitive}. More recently, a high-quality resonator with low parasitic capacitance was combined with varactors to improve significantly the charge sensitivity in the range of 55\,mK to 1.5\,K \cite{ibberson2019low}. Such a limitation in the use of rf reflectometry can be severe in certains architectures, like in p-channel silicon devices where mobilities are lower compared to the ones found in n-type devices. This generates an excess resistance in the 2DHG and decreases the readout sensitivity rendering rf measurements more challenging. In this study, we purposely use p-channel Metal-Oxide-Semiconductor (PMOS) devices to demonstrate such an rf leakage effect induced by the top gate and discuss the ability to perform rf reflectometry.
	
	\section*{Results}
	
\begin{figure}[t]
		\includegraphics[width=1\columnwidth]{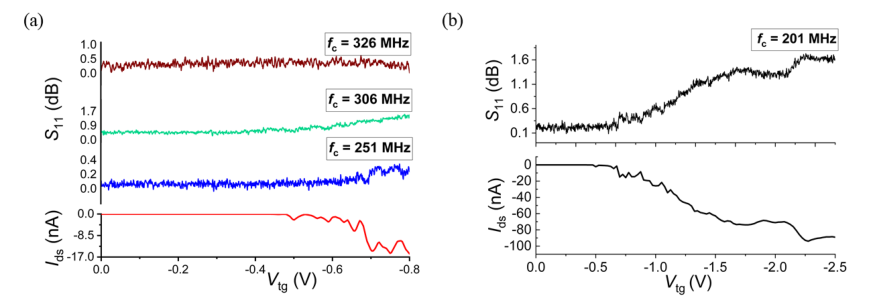}
		\caption{The rf-response at various probing frequencies for device 2.(a) Reflectivity response $S_{\textup{11}}$ together with the corresponding DC current as a function of the top gate voltage $V_{\textup{tg}}$ at 4.2\,K. (b) Comparison of the rf- and DC responses in the accumulation, depletion and inversion regions.}\label{fig:cutoff}
	\end{figure}

	We first used device 2 and determined the parasitic capacitance $C_{\textup{p}}$ of the tank circuit by monitoring the variation of the resonance frequency $f_{\textup{r}} = 1/2\pi \sqrt{L C_{\textup{p}}}$ with the inductance value $L$. We obtained $C_{\textup{p}} \sim$ 0.48 pF. Several inductors were then used to obtain a set of resonant frequencies $f_{\textup{r}}$ between 401\,MHz and 201\, MHz that we used as carrier frequencies. For each inductors, the top gate voltage $V_{\textup{tg}}$ was swept and both the DC conductivity and the reflective RF signal were measured as shown in Fig. \ref{fig:cutoff}. We noticed that for \textcolor{black}{probing frequency} $f_{c}\gg$ 326 MHz and despite having a clear DC signal, no reflective signal was observed. This is a consequence of the too high value of the parasitic capacitance and consequently a gate impedance much lower than the matching circuit impedance. However, an rf-response is clearly observed for frequencies lower than 306\,MHz. One can find out that the impedance of the parasitic capacitance decreases as the probing frequency increases allowing determining an upper bound for the usable probing frequency in our rf-setup. We should underline that all measurements have been performed under relatively good matching conditions with the depth of resonant dip exceeding 7\,dB, which indicates that the impedance matching mostly affects the measurement sensitivity and not the ability to perform reflectivity measurements itself.

	Despite the parasitic capacitance in device 2 being reduced compared with even larger devices ($C_{\textup{p}} \approx$ 0.6\,pF for a device with a top gate of 4$\mbox{$\times$10}^2$ $\mbox{$\mu$m}^2$), this still remains an issue for measurements requiring a high sensitivity. To mitigate this, we used the smallest gate area device (device 1) for which $C_{\textup{p}} \sim$ 0.26 pF for the remaining experiments. \textcolor{black}{It is worth mentioning that several parameters affect the parasitic capacitance, such as the material used in the printed circuit board (PCB), the distance between the PCB and the device, as well as the number of bonding wires used to connect the device to the PCB. Moreover, capacitance components are distributed in actual CMOS devices. Therefore, there is not necessarily a direct ratio between the reduced area of the top gate and parasitic capacitance obtained.}
	
\begin{figure}[t]
		\includegraphics[width=1\columnwidth]{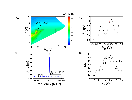}
		\caption{SNR measurement of device 1. (a) Variation of the reflection's magnitude as a function of varactor voltages $V_{\textup{m}}$ vs $V_{\textup{t}}$. \textcolor{black}{ $\Delta$ $S_{\textup{11}}$ represents the amplitude in dB while voltages are applied to both varactors. The minimum amplitude is 9 dB and, it is obtained in the case $V_{\textup{m}} =$ 0\,V, $V_{\textup{t}} =$ 10\,V.} Best matching is obtained for $V_{\textup{m}}$ = 1.5\,V ($\sim$ 5.5\,pF) and $V_{\textup{t}}$ = 5\,V ($\sim$ 6\,pF) giving a maximum reflective depth of 63\,dB for $\Delta$ $S_{\textup{11}}$
			(b) Gate oscillations for device 1 as a function of top gate voltage \textcolor{black}{($V_{\textup{tg}}$)}  at 4.2\,K with $V_{\textup{t}} = 5$\,V and $V_{\textup{m}} = 1$\,V. The operating point was set to the maximum transconductance $V_{\textup{tg}}=$ -1.85\,V for the SNR measurements in order to achieve the maximum sensitivity. (c) Modulation spectrum showing the 3\,kHz sidebands due to the modulating signal ($f_{\textup{m}}$) of amplitude 10\,mV rms applied to the top gate. (d) Variation of the SNR with $V_{\textup{m}}$ at $V_{\textup{t}} =$ 5\,V showing a maximum value at $V_{\textup{m}}$ = 1.5\,V. }\label{fig:Fig4}
	\end{figure}

	We first check the ability to independently adjust the  $f_{\textup{r}}$ and the sensibility of the reflectometry circuit by mapping out the variation of the reflection coefficient $S_{\textup{11}}$ in terms of $V_{\textup{t}}$ and $V_{\textup{m}}$, \textcolor{black}{associated with varactor capacitance $C_{\textup{t}}$ and $C_{\textup{m}}$, respectively,} over the range 0 to 10 V. As expected the resonance amplitude, and so the quality factor $Q$ of the matching circuit, changes with $V_{\textup{m}}$ without significantly affecting the resonant frequency value $f_{\textup{r}}$ \cite{ares2016sensitive}. We then selected a reduced range of varactor voltages and optimised the $S_{\textup{11}}$ parameter to obtain the best matching condition among our measurements (see S2). We observe two set of possible voltage values ($V_{\textup{t}}$, $V_{\textup{m}}$) for which the $\Delta S_{\textup{11}}$ value is maximum (Fig. \ref{fig:Fig4}(a)). Despite $V_{\textup{t}}$ being able to improve the matching, it mostly controls $f_{\textup{r}}$ due to its proximity to the circuit. On the contrary, $C_{\textup{m}}$ is located upfront of the matching input point, making it more suitable and more efficient for matching purposes. Best matching is obtained for $V_{\textup{m}}$ = 1.5 V and $V_{\textup{t}}$ = 5 V, corresponding to $C_{\textup{t}} = 6$\,pF and $C_{\textup{m}} = 5.5$\,pF. This gives a reflective signal depth of 63\,dB. In the view of increasing the measurement sensitivity, namely the quality factor, a higher value for the inductor would be needed but this would ultimately reduce the bandwidth and introduce unintentionally inductor self-resonances around the carrier frequency.  
	
	In order to correlate the depth of the impedance matching to the SNR, we set $V_{\textup{t}} = 5$\,V and adjusted the value of the top gate to the maximum of the transconductance, i.e. -1.85\,V (Fig. \ref{fig:Fig4}(b)). We then applied a small ac-signal of amplitude 10\,mV rms and frequency 3\,kHz on the top gate and observed the amplitude modulation (AM) spectrum of the quantum dot prior demodulation. The visibility of the sidebands at $f_{\textup{c}}\pm$\,3\,kHz allows determining the SNR value (Fig. \ref{fig:Fig4}(c)). We were then able to obtain the variation of the measurement sensitivity as a function of $V_{\textup{m}}$ while keeping the resonant frequency unaffected.
	
\begin{figure}[t]
		\includegraphics[width=1\columnwidth]{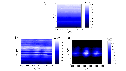}
		\caption{Charge stability diagram of device 1 obtained by rf reflectometry for three different impedance matching conditions. (a) Both varactors are turned off, (b) Impedance matching is away from best matching ($V_{\textup{m}}$ = 1.1\,V, $V_{\textup{t}}$ = 5\,V), (c) Best matching condition ($V_{\textup{m}}$ = 1.5\,V, $V_{\textup{t}}$ = 5\,V)}\label{fig:chargeStability}
	\end{figure}

	As observed previously, the maximum SNR is obtained at $V_{\textup{m}} = 1.5$\,V implying a best matching condition (Fig. \ref{fig:Fig4}(d)). However, we did perform additional rf measurements at different matching conditions to monitor the value of the sensitivity. In the case both varactors were turned off, we did not observed any Coulomb diamonds (Fig. \ref{fig:chargeStability}(a)). Away from optimal matching, Coulomb diamonds were barely discernable (Fig. \ref{fig:chargeStability}(b)) while their visibility was largely improved under best matching (Fig. \ref{fig:chargeStability}(c)). These results confirm the increase in SNR as the matching conditions get optimised. 
	
	\section*{Discussion}

	The charge sensitivity is given by $\delta_q = \Delta_{\textup{q}}$ $_\textup{rms} /{(\sqrt{2\textup{RBW}}. 10^{\textup{SNR}/20}})$ where  $\Delta$$\textup{q}$$_\textup{rms}$ is the applied root-mean-square gate charge, RBW the resolution bandwidth, and SNR the signal-to-noise ratio. At the best matching point, we obtained $ \Delta$$\textup{q}$$_\textup{rms}$ $\sim 0.11e_{\textup{rms}}$, RBW $=5.1$ Hz, and SNR $=15.5$ dB giving $\delta_q \approx{6.10^{-3}} e/\sqrt{\textup{Hz}}$.
	The maximum sensitivity in pure rf mode has been calculated by Korotkov and Paalanen \cite{korotkov1999charge}. In our setup, the sensitivity is mainly limited by the tunnel resistance, $R_\textup{t} = 5.10^7\  \Omega$, and total island capacitance, $C_\Sigma =10^{-17} \ \textup{F} $ which are both device related. 
	Having a relatively high resistance in PMOS devices results in a low charge sensitivity.
	Other limiting factors include the noise generated by room temperature amplifiers, the losses at the inductor, and dielectric losses at the PCB. All account for the decrease in sensitivity. \textcolor{black}{ We should mention that thermal noise and device characteristics may be responsible of irregularity in the Coulomb diamonds even at the best matching condition (Fig. \ref{fig:chargeStability}(c)).}
	
	As a conclusion, we have determined an upper bound for the usable frequency in our rf-reflectometry setup above which the gate impedance becomes sufficiently small and the rf-leakage to the ground gets dominant. By tuning the frequency into the appropriate range and by using two GaAs-based varactors to alleviate the stray capacitance, tune the resonant frequency and improve the matching condition of the circuit, we were able to perform an rf-readout of physically defined p-channel silicon quantum dots. Under best matching condition and optimum SNR, we successfully observed Coulomb diamonds at 4.2\,K. This study underlines the necessity of carefully designing the device structure in order to achieve high measurement sensitivity and increase the operability of the rf-reflectometry. The addition of the two varactors in the circuit allows the necessary tuning flexibility which will be required in a multi-qubit platform, where crosstalks are unavoidable and frequency multiplexing is expected to be used for both addressing and readout. The demonstration of optimised circuitry at 4.2\,K is promising for future high temperature operation of quantum bits.
	
	\section*{Methods}
	\subsection*{Device fabrication}
\begin{figure}[t]
		\includegraphics[width=1\columnwidth]{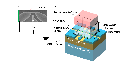}
		\caption{Device structure. (a) Scanning electron microscope (SEM) image of the device after reactive ion etching. Darkest areas are parts of the SOI that have been etched away whereas the island at the centre is the quantum dot under study with its associated control gate $G_{\textup{SDQ}}$, (b) Device internal structure once fully processed.}\label{fig:device}
	\end{figure}

	Scanning electron microscope (SEM) image and schematic image of the device are shown in Fig. \ref{fig:device}(a) and \ref{fig:device}(b), respectively. All devices were fabricated from an undoped 40-nm-thick silicon-on-insulator (SOI) with a 145-nm-thick buried oxide. \textcolor{black}{Quantum dot structures and the side gate were then formed by etching the silicon active layer at selected regions using inductive coupled plasma reactive ion etching (ICP-RIE) \cite{kambara2013dual,horibe2015back,yamaoka2017charge,bugu2021rf}. However, in these experiments the side gate was not used.} After a standard RCA clean, a 3-nm-thick thermal oxide layer was grown with dry oxidation at 850$^\circ$ before the deposition of a 65-nm-thick gate oxide. A top gate made from highly phosphorus-doped polysilicon at a concentration of $10^{-20}$ cm$^{-3} $ was then deposited. Two devices are used in our study, both having the same structure but different top gate areas, one with a surface area of 0.09\,$\mbox{$\mu$m}^2$ (device 1) and the other with an area of 1\,$\mbox{$\mu$m}^2$ (device 2).
	
	\subsection*{Measurement setup}
\begin{figure}[t]
		\includegraphics[width=0.8\columnwidth]{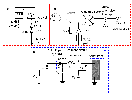}
		\caption{Measurement setup. (a) Room temperature part of the circuit used for the resonant frequency $f_{\textup{r}}$ dependence measurement. The rf signal at 0\,dBm is first attenuated before entering a directional coupler and being sent to the quantum dot via the resonant circuit at low temperature. The reflected signal is then amplified at high gain before being measured at the port 2 of the vector network analyzer (VNA). (b) Room temperature part of the circuit used to readout the charge state. Here, the signal from the rf-source is split between the incident signal sent directly to the device at low temperature and the demodulator for the measurement of the in-phase (I) and quadrature (Q) components of the reflected signal. (c) The resonant circuit with an inductor of 277\,nH together with the two varactors $C_{\textup{m}}$ and $C_{\textup{t}}$. $R_\textup{d}$ represents the losses in the measurement system including  the dielectric losses in the device, the printed circuit board (PCB), and in tuning the varactors.}\label{fig:Setup}
	\end{figure}

	Experiments are performed using the setup shown in Fig. \ref{fig:Setup}. Figures \ref{fig:Setup}(a) and (b) refer to the room temperature measurement setup while Fig. \ref{fig:Setup}(c) shows the cryogenic part of the setup used at 4.2 K. The rf-signal passing through the directional coupler is first attenuated with a -40 dB attenuator before be sent to the quantum dot and the tank circuit. The reflected signal is then amplified with room temperature amplifiers with a total gain of 60 dB either using  the setup (a) or (b) depending on the measurement intended. For the resonant circuit, we used a GaAs double varactor (a Macom MA46H202-1056 for $C_{\textup{m}}$ and a Macom MA46H204-1056 for $C_{\textup{t}}$) in order to tune both the resonance frequency to the region of operation and the impedance matching. This allows reading out the charge states of a PMOS quantum dot at 4.2 K, under high SNR and up tp perfect the matching conditions. For the measurement of device 1, we added a matching capacitor $C_{\textup{m}}$ as well as a decoupling capacitor $C_{\textup{d}}$ ($\sim$ 24 pF) at the input point of the matching circuit. The aim of $C_{\textup{d}}$ is to increase the quality factor $Q$ of the circuit by decoupling it from the input. The varactor $C_{\textup{t}}$ controlled by $V_{\textup{t}}$, aims at controlling the resonant frequency $f_{\textup{r}}$ and the loaded impedance $Z_{\textup{L}}$ while the second varactor $C_{\textup{m}}$, controlled by $V_{\textup{m}}$, mostly adjusts the value of $Z_{\textup{L}}$ and hence the sensitivity (see S2). The rf-signal was added via a bias tee to the drain of the device with all the electronic components mounted on a standard FR4 printed circuit board (see S1).
	
	In rf experiments, the choice for $f_{\textup{c}}$ is dictated by both noise and circuit impedance considerations, typically. To maintain a high SNR, $f_{\textup{c}}$ should always exceed the $1/f$ noise caused by charge fluctuations \cite{Schoelkopf_1998}. On the other hand, if the rf-signal is connected to the leads, the impedance of the matching circuit $Z_{\textup{m}}$ shall not be much larger than the gate impedance $Z_{\textup{g}} = 1/2\pi \dot{\imath} f_{\textup{c}} C_{\textup{g}}$ where $C_{\textup{g}}$ is the parasitic capacitance of the gate, if one wants to avoid leakage of the rf-signal. Such a leakage effect is well demonstrated in the device with the larger gate area (device 2) by exploring its rf-response at different carrier frequencies and so, at different values of the resonant circuit inductance.
	
	\bibliography{ms}
	

	\section*{Acknowledgements}
	
	S. B. would like to thank M. F. Gonzalez-Zalba, J. Yoneda, and H. Takahashi for helpful discussion. S. B. also thanks the Hitachi Cambridge Laboratory for the hospitality. This work was partly supported by JST CREST (JPMJ CR1675), JSPS KAKENHI (Grant Numbers JP18K18996 and JP20H00237), MEXT Quantum Leap Flagship Program (Q-LEAP) Grant Number JPMXS0118069228, JST Moonshot R\&D Grant Number JPMJMS2065 as well as MEXT Project for Developing Innovation S\texttt{}ystems.

	\section*{Author contributions statement}
	
	S.B. designed the scheme, and conducted the experiments. S.B. also provided feedback for device fabrication. S.N., K.K., Y.L., S.M., T.M., and T.K. fabricated the devices. S.B., T.F. and T.K. analyzed the results and wrote the manuscript. T.K. managed the project. 
	
	

	
\end{document}